# Broadband laser polarization control with aligned carbon nanotubes

He Yang[a†], Bo Fu[a,b†], Diao Li[a,c†], Ya Chen[a], Marco Mattila[a], Ying Tian[d], Zhenzhong Yong[e], Changxi Yang[b], Ilkka Tittonen[a], Zhaoyu Ren[c], Jingtao Bai[c], Qingwen Li[e], Esko I. Kauppinen[d], Harri Lipsanen[a], and Zhipei Sun*[a]

We introduce a simple approach to fabricate aligned carbon nanotube (ACNT) device for broadband polarization control in fiber laser systems. The ACNT device was fabricated by pulling from as-fabricated vertically-aligned carbon nanotube arrays. Their anisotropic property is confirmed with optical and scanning electron microscopy, and with polarized Raman and absorption spectroscopy. The device was then integrated into fiber laser systems (at two technologically important wavelengths of 1 and 1.5 μm) for polarization control. We obtained a linearly-polarized light output with the maximum extinction ratio of ~12 dB. The output polarization direction could be fully controlled by the ACNT alignment direction in both lasers. To the best of our knowledge, this is the first time that ACNT device is applied to polarization control in laser systems. Our results exhibit that the ACNT device is a simple, low-cost, and broadband polarizer to control laser polarization dynamics, for various photonic applications (such as material processing, polarization diversity detection in communications), where the linear polarization control is necessary.

## 1. Introduction

Carbon nanotubes (CNTs) [1] have been extensively investigated due to their inherent physical properties (e.g., electrical,[2] optical,[3] and thermal [4,5] properties), enabling various applications, such as electron field emitters,[6] quantum resistors,[7] transistors,[8] atomic force microscopes,[9] mechanical memory elements,[10] etc.[11-15] Worth noting that, CNTs have recently attracted huge attention for various photonic and optoelectronic applications[2] because of their unique optical properties, such as broadband optical absorption,[16-18] large optical nonlinearity,[19] ultrafast carrier relaxation time,[20] and high damage threshold.[21] One of the most successful examples is CNT-based saturable absorber, [22,23] which has been employed for ultrafast pulse generation in solid-state,[24] semiconductor,[25] waveguide,[26] and fiber lasers.[27]

Due to the unique structure confinement in one-dimension, aligned CNTs (ACNTs) have been demonstrated to facilitate a large range of applications (such as touch screen,[28] solar cells,[29] sensors,[30] supercapacitors,[31] thermal interface materials,[32] batteries,[33] and THz sources[34]), underscoring superior performance compared to their random oriented counterpart. In particular, ACNTs have strong optical anisotropic characteristic,[35,36] enabling polarizer applications with superior performance when compared to the conventional bulk polarizers (e.g., prisms), such as easy fabrication, broad operation bandwidth (from ultraviolet,[18] visible,[37] infrared,[38] to THz range[39,40]), and high extinction ratio (up to 30 dB [38]).

Laser polarization control is very crucial for a large range of photonic applications, ranging from fluorescence imaging, liquid crystal device characterization and manufacturing, to laser material processing and polarization diversity detection in communications and range finding. In this paper, we introduce a simple technique to fabricate ACNT device, by which broadband (at 1 and 1.5 μm) laser polarization dynamics is fully controlled after integration into fiber laser systems. The maximum extinction ratio of ~12 dB is achieved. To the best of our knowledge, it is the first time that such ACNT device is applied to the fiber laser systems for polarization control. Our study demonstrates the unique broadband performance of ACNTs for various photonics and optoelectronics applications.

## 2. Fabrication and characterization of the ACNT device

The ACNT device was made by pulling out from a vertically-aligned CNT array,[28] which was fabricated with chemical vapour deposition method.[41] Then the ACNT film was directly transferred onto the surface of a quartz substrate (as shown in Fig. 1 (a) after inserting into a mirror mount). Note that our direct-dry transfer fabrication method is also compatible to other photonic devices (e.g., fibers and silicon devices).

In order to characterize the alignment performance, various microscopies were utilized, such as optical microscopy (Fig. 1(b)), scanning electron microscopy (SEM, Fig. 1 (c) and (d) with different scales). As shown in the figure, CNT arrays were highly-aligned along the pulling direction. The diameter of the CNTs was measured to be ~10 nm by SEM, which was also confirmed by Transmission Electron Microscopy (TEM). It denotes that our sample consists typically of multi-wall CNTs.

[a] Department of Micro- and Nanosciences, Aalto University, PO Box 13500, FI-00076 Aalto, Finland. E-mail: zhipei.sun@aalto.fi; Tel: +358-50-4322979

[b] The State Key Laboratory of Precision Measurement Technology and Instruments, Department of Precision Instruments, Tsinghua University, Beijing 100084, China

[c] Institute of Photonics and Photo-Technology, School of Physics, Northwest University, Xi'an, Shaanxi 710069, China

[d] Department of Applied Physics, Aalto University School of Science, PO Box 15100, FI-00076 Aalto, Finland

[e] Institute of Nano-tech and Nano-bionics, Chinese Academy of Sciences, Suzhou, Jiangsu 215125, China

† He Yang, Bo Fu and Diao Li contributed equally to this work.

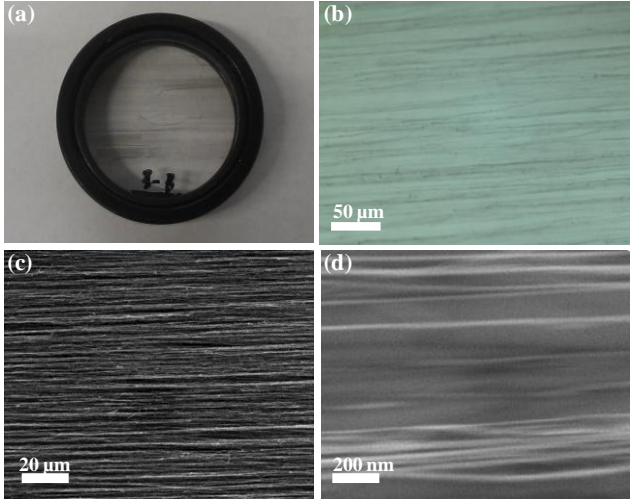

Fig. 1 (a) ACNT on 1-inch quartz substrate, (b) with its optical microscope image, (c) and (d) SEM images with different scales.

For ACNTs, their Raman scattering intensity varies with different excitation polarization direction. In our experiment, a polarized Raman spectroscopy (WITec Alpha 300 RA) was utilized to characterize the anisotropic properties of the ACNT samples. Polarized Raman spectra were conducted by changing the input polarization direction ($\phi$) of the excitation laser from 0° to 360° with respect to the CNT alignment direction. When $\phi$=0° and $\phi$=90°, the incident excitation polarization direction is parallel and perpendicular to the ACNT alignment direction, respectively. Typical Raman spectra (Fig. 2 (a)) showed three dominating features, namely D, G, and G' modes, as typically observed for multi-wall carbon nanotubes.[42, 43] It has been reported that the G-band of Raman modes at around 1580 cm$^{-1}$ is highly sensitive to CNTs alignment.[42] When we change the angle $\phi$, the intensity of G-peak and D-peak changes from the maximum (at $\phi$=0°) to the minimum (at $\phi$=90°), which was in accordance with the polarized absorbance of the aligned carbon nanotube.[44,45] The ratio between the maximum and the minimum intensity is ~22. In our study, the G-peak intensity at different angle $\phi$ is presented in Fig. 2 (b). It indicates that the Raman scattering intensity of ACNTs is sensitive to the polarization direction of the pump light, well agreeing with the equation of $I(\phi) \propto \cos^2(\phi)$ [42] for ACNTs. Such anisotropic Raman property clearly shows that CNTs are well aligned.

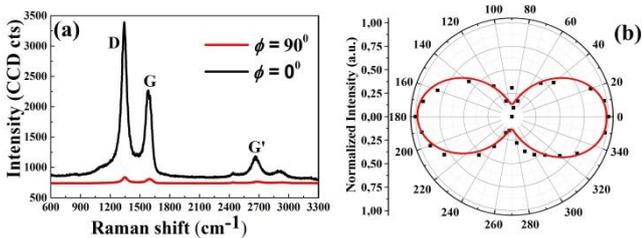

Fig. 2 (a) Polarized Raman spectra of the ACNT device, (b) G-peak Raman intensity as a function of the angle, and the red line depicts the fitting result with the equation. The excitation laser wavelength is 532 nm.

Fig. 3 (a) shows the polarized transmittance of our ACNT device in the two orthogonal directions (i.e., when the polarization direction of the input light is parallel or perpendicular to the CNT alignment direction), which is measured by a polarized absorption spectroscopy. When the incident light polarization is perpendicular to the ACNT alignment direction, the transmittance is ~94% at 1.8 µm, which is mainly contributed by Fresnel loss (~6%) of the quartz substrate. While for the parallel polarization light input, the transmittance is ~82% at 1.8 µm. The result identifies the optical anisotropic absorption of our CNT device. The difference between two orthogonal directions is ~12% at 1.8 µm (~16% at 300 nm, as shown in Fig. 3 (b)), which is comparable to the typical performance reported for aligned carbon nanotubes (from 15% to 20% at different wavelengths [39,40]). Note that the absorption difference remains almost constantly from 1 to 2 µm, which shows the unique broad operation bandwidth of our device.

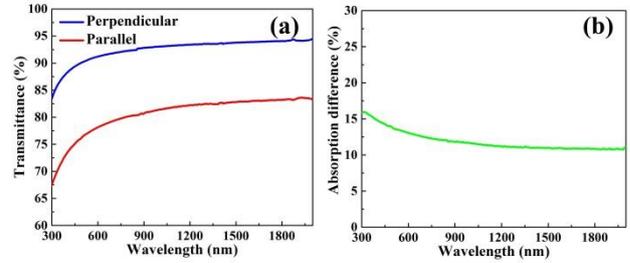

Fig. 3 (a) Polarized transmittance of the ACNT device with the aligned direction parallel and perpendicular to the polarization direction of the incident light source, respectively. (b) The absorption difference in the two orthogonal directions.

## 3. Laser setup

To utilize the broadband anisotropic characteristics of the ACNT device, fiber lasers operating at 1 and 1.5 µm were designed with identical layout (Fig. 4). In the fiber laser at 1.5 µm: a 1 m erbium doped (Er-doped) fiber worked as the gain fiber, a wavelength division multiplexer (WDM) was used to couple the pump laser (980 nm laser diode (LD)) into the gain fiber, the ACNT device was inserted inside a fiber-collimator based U-bench to modulate the intra-cavity polarization state, a high-reflectivity fiber based mirror (made by a 3-dB coupler as a nonlinear optical loop mirror) was provided for laser feedback, while a fiber coupler with 20% coupling ratio outputs the laser from the cavity for measurement. In the 1-µm fiber laser: the gain fiber was replaced by 0.5 m ytterbium doped (Yb-doped) fiber, and other components (e.g., WDM, LD, and coupler) with same function were selected at 1 µm. In both laser systems, all fiber components were polarization independent, indicating no intra-cavity polarization preference without ACNTs.

An optical spectrum analyzer (Hewlett Packard, 86140A) was utilized to measure the laser output spectrum. A polarization analyzer (i.e., a linear polarizer) mounted in a high-precision rotator was fixed after the output coupler. By rotating the polarization analyzer from 0° to 360°, the output laser power after the analyzer was monitored by a power meter (OPHIR Nova II) to characterize the output polarization state.

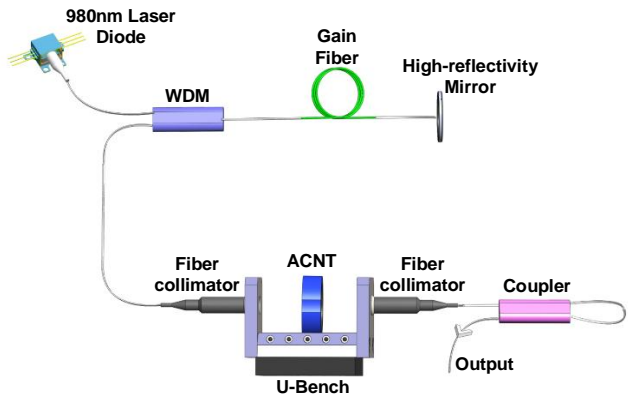

Fig. 4 Fiber laser setup for both 1 and 1.5 μm laser systems.

# 4. Polarization control results and discussions

## 4.1 1.5 μm Er-doped fiber laser (EDFL)

In EDFL, continuous wave output performance was studied with and without the ACNT device inserted in the U-bench (Fig. 4). Fig. 5 (a) shows the typical output optical spectrum, identical for both cases (with and without the ACNT device). The peak wavelength is ~1561 nm with the full width at half maximum (FWHM) of 0.1 nm, a typical value for such erbium-doped fiber laser without mode-selection. The output power as a function of the pump power is shown in Fig. 5 (b). The output power presents linear characteristics for both situations, while the slope efficiency decreased when ACNT was inserted. It is because the ACNT device in the cavity also introduces small attenuation (~6% as shown in Fig. 3 (a)) to the laser cavity. The laser threshold pump power was ~34 mW. The maximum output power decreased from ~1.27 to ~1.1 mW under the highest pump power of ~180 mW (the maximum output power from the laser diode), after inserting the ACNT device in the cavity. Further increase of the laser output power is possible, as the output power linearly increases with the pump power with no sign of saturation.

To investigate the polarization dynamics of our ACNT device in the laser resonator, we experimentally measured the laser output power after the polarization analyzer. At first, we measured the power change without inserting the ACNT device (the inset in Fig. 5(d)). The output power is almost invariable with a small power fluctuation (~1.9%). It indicates that the output polarization of the laser without the ACNT device is random (i.e., not linearly polarized), which is expected due to non-polarization preference in our polarization-independent fiber laser cavity. The small power fluctuation is most likely due to the environment perturbation (e.g., temperature or vibration). After that, we inserted the ACNT device into the laser resonator and fixed the alignment direction of ACNT vertical to the horizontal plane in the free space, without changing any experimental conditions. Similarly, the output laser power after the polarization analyzer was recorded by rotating the polarization analyzer from 0° to 360°. The experimental results are illustrated in Fig. 5(c), fitted well with the cosine function, i.e., the minimum output power was close to zero at the angle of 0°, while the maximum power located at the angle of 90°. It shows that the output polarization of the laser was changed from the random-polarization state to the linear-polarization state, after insertion of our ACNT device in the laser cavity.

In order to fully study the performance of intra-cavity polarization control of ACNT device, the CNT alignment direction was rotated from vertical to horizontal orientation for comparison. The corresponding results are also shown in Fig. 5-(c). It is observed that the whole curve shifts 90° when comparing to the one with vertical orientation of the ACNT device. The result suggests that the laser output with ACNT device was linearly-polarized, and the polarization direction rotates with the rotation of the ACNT device. This shows that the polarization state of our laser output can be simply controlled by rotating the intra-cavity ACNT device.

Fig. 5(d) summarizes the degree of polarization (DOP) and extinction ratio (ER) of our ACNT laser versus pump power. The DOP and ER are defined as DOP=$(P_{max}-P_{min})/(P_{max}+P_{min})$, ER=$P_{max}/P_{min}$, where $P_{max}$ and $P_{min}$ are the maximum and minimum output power given by the measured results shown in Fig. 5 (c). The maximum DOP of our ACNT laser is 87.5% at the pump power of ~75 mW, 10-time higher than the DOP of the laser without ACNT (~8%). The corresponding ER of our ACNT laser is ~12 dB, around 8-time higher than the value of the laser without ACNT (~1.5 dB). The ER is ~9.8 dB (with 81% of DOP) at the maximum output power of 1.1 mW. The DOP and ER performance is pretty unchanging when the pump power is high (>100 mW). This is because the anisotropic absorption of ACNT dominates the laser output polarization performance at the high pump power.

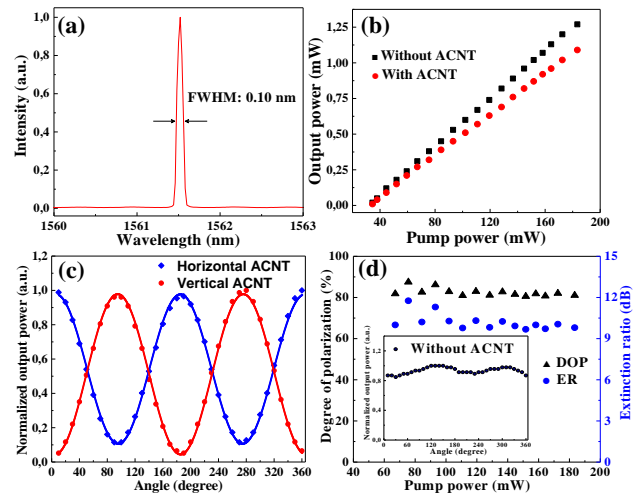

Fig. 5 Experimental results at 1.5 μm: (a) output spectrum (b) output power as a function of pump power (with and without ACNT), (c) normalized output power as a function of polarization analyser angle of our ACNT laser output, and the cosine function fitting of the experimental data (d) DOP and ER versus pump power, and the inset figure is the normalized output power as a function of polarization analyser angle without ACNT insertion.

## 4.2 1 μm Yb-doped fiber laser (YDFL)

In our ACNT based YDFL, the output spectrum is shown in Fig. 6 (a). The peak wavelength is ~1032 nm with FWHM of ~0.1 nm, typical for an YDFL without any wavelength selection component in the cavity. The spectrum is similar to the laser without the ACNT device. Fig. 6 (b) shows the results of laser output power with and without ACNT device. The threshold for the laser system with ACNT (21 mW) was slightly higher than that (20 mW) without ACNT, because of the small attenuation introduced by the ACNT device. With the increase of the pump power, the output power increased linearly. The highest output power for the laser without ACNT is ~2.3 mW, while for the laser with ACNT, the output power is ~2.1 mW. The difference is also due to the small attenuation of our ACNT device.

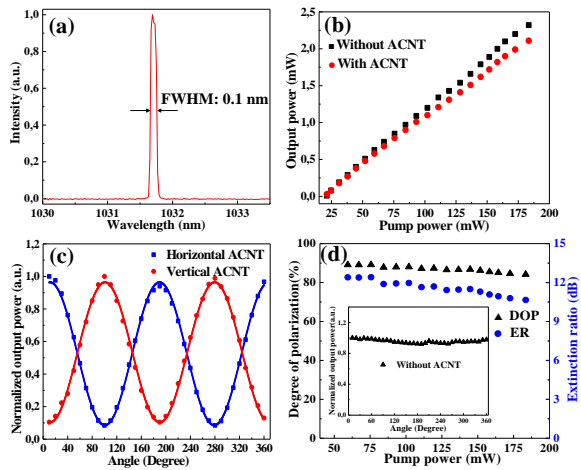

Fig. 6 Experimental results at 1 μm: (a) output spectrum, (b) output power as a function of pump power, (c) normalized output power as a function of angle with the polarization analyser, and the cosine function fitting of the experimental data, (d) DOP and ER versus pump power, and the inset figure is the normalized output power as a function of polarization analyser angle when the laser is without the ACNT device.

Fig. 6 (c) presents the polarization control results of our 1-μm YDFL with ACNT device. Similar to the EDFL experiment, we also measured the laser output power change after the polarization analyzer without inserting the ACNT device, as illustrated in the inset figure of Fig. 6(d). The output power keeps almost unvarying with the fluctuation of ~5.8%, which indicates that the output of our laser without the ACNT is randomly-polarized, similar to what we observed with the laser working at 1.5 μm. Then we inserted the ACNT device into the resonator, and used the experimental procedures (identical to what we employed in EDFL) to study the polarization state of the laser output. The result is given in Fig. 6(c). As expected, the output of our laser with ACNT became linearly polarized, and the polarization state can be controlled by the rotation of the ACNT alignment direction. Fig. 6 (d) summarizes the DOP and ER versus pump power. The DOP reaches ~89% at the pump power of 59.4 mW (in contrast to the DOP of 2.9% without the ACNT device in the laser cavity). It decreased slightly as we increased the pump power, which also confirms good polarization dynamics control of our 1 μm fiber laser with the ACNT device. The maximum ER is ~ 12.4 dB, comparable to the result achieved with the EDFL.

## Conclusions

We introduce a simple method to fabricate ACNT device for broadband polarization control in laser systems (at 1 and 1.5 μm). The anisotropic property of ACNTs was confirmed with various characterization methods (e.g., optical microscopy, SEM, and Raman spectroscopy). By integrating the ACNT device into the filer laser system, we obtained a linearly-polarized light output with DOP up to 89.1% and 87.5% with corresponding ER of 12.4 and ~12 dB in 1 and 1.5 μm laser systems, respectively. Our experimental results exhibit that the ACNT device can be potentially employed as polarization controller for a broad range of photonic applications (nonlinear frequency conversion,[46,47] beam combination,[48] material processing and polarization diversity detection in communications etc.), where the linear polarization output is required.

## Acknowledgements

The authors acknowledge funding from Teknologiateollisuus TT-100, Academy of Finland (grants: 276160, 276376, 284548, 285972), the European Union's Seventh Framework Programme (REA grant agreement No. 631610), National Natural Science Foundation of China, China Scholarship Council, and Aalto University (Finland). The authors also thank the provision of facilities and technical support by Aalto University at Micronova Nanofabrication Centre.

**Author contribution:**

Z.S., H.Y., B.F. and D.L. conceived and designed the laser experiments. Z.S., Y.C., M.M., Y. T. performed the characterization experiments. Z.Y. and Q.L. fabricated the ACNT device. I.T., Z.R., J.B., Q.L., E.K., H.L., and Z. S coordinated the experiments. All authors contributed to the writing of the manuscript and to the discussion.

## Notes and references